\documentclass[runningheads,a4paper]{llncs}
\usepackage[utf8]{inputenc} 
\usepackage[T1]{fontenc}
\usepackage{amssymb}
\usepackage{graphicx}
\usepackage{listings}
\DeclareGraphicsExtensions{.eps,.jpg}
\usepackage{url}
\newcommand{\keywords}[1]{\par\addvspace\baselineskip
\noindent\keywordname\enspace\ignorespaces#1}

\begin{document}

\mainmatter

\title{Towards Autotuning of OpenMP Applications on Multicore Architectures}  
\titlerunning{Towards Autotuning of OpenMP Applications}

\author{Jakub Katarzy\'nski \and Maciej Cytowski}

\institute{Interdisciplinary Centre for Mathematical and Computational Modelling,\\ University of Warsaw}

\maketitle

\begin{abstract}

In this paper we describe an autotuning tool for optimization of OpenMP applications on highly multicore and multithreaded architectures. Our work was motivated by in-depth performance analysis of scientific applications and synthetic benchmarks on IBM Power 775 architecture. The tool provides an automatic code instrumentation of OpenMP parallel regions. Based on measurement of chosen hardware performance counters the tool decides on the number of parallel threads that should be used for execution of chosen code fragments.    

 
\end{abstract}


\section{Introduction}

Performance of today's general purpose processor architectures is driven by three main components: clock speed, number of computational cores and number of operations per clock cycle. Since further increasing the clock speed remains very difficult or even impossible, the performance improvement over previous generations has to come from the latter two factors. In particular, hardware vendors tend to develop different ways of increasing single core performance i.e. vector processing, fused multiply and add operations and support for simultaneous processing of multiple threads. On the other hand, the largest HPC systems nowadays are equipped with millions of cores, requiring the use of mixed process and thread based parallelism. As a result, applications that will run on future HPC systems will need to demonstrate very high thread based scalability within the compute nodes. This task is non-trivial for many complex applications performing different computational algorithms. In this paper, we describe our methodology for performance analysis and autotuning of OpenMP applications on multi-core and multi-threaded nodes of IBM Power775 (IH) system. The system itself is described in Section \ref{platform}

Our work is motivated by measurements that we have performed on IBM Power 775 system with scientific applications and synthetic benchmarks which are presented in Section \ref{motivation}. 

Firstly, we are very keen to know why performance of chosen algorithms can benefit from using multithreading while others do not. We believe that this problem might be addressed by detailed analysis of hardware performance counters. For instance we are
analyzing hardware performance counters for the chosen algorithms from the BOTS benchmark. We are looking for a
correlation between the ability to efficiently use the multithreading mechanism and the performance profile of the given application. Such a result would lead us to better understanding of computational nature of different algorithms, but it could also be used to propose an automatic heuristic algorithm (e.g. based on decision trees) to decide the number of threads used by specific code fragments during execution.

Secondly, many modern HPC applications use both MPI and thread parallel model (e.g. mixed MPI + OpenMP). Parallel
processes executed on different computational nodes include many thread parallel regions which are executed on available
computational cores. Very often the number of threads in thread based parallelization is controlled by a single switch (e.g. the OMP\_NUM\_THREADS environment variable). Since different algorithms and code fragments may present different scalability, it would be appropriate to choose number of threads for execution to each parallel region individually.
Moreover such a decision could be made automatically only with a minor information from the user.

In Section \ref{tools} we describe existing tools together with design and implementation of our new tool. We show the differences and describe the usage scenario.

\section{Computing platform}
\label{platform}

Reasearch and implementation presented in this paper was carried out on the IBM Power 775 (IH) supercomputing server - a highly scalable system with extreme parallel processing performance and dense, modular packaging. It is based on IBM Power7 architecture and was designed by IBM
within the US DARPA's HPCS (High Productivity Computing Systems) Program. This unique supercomputing
environment is currently available in few organisations worldwide, e.g.: ICM, University of Warsaw (Poland),
Met Office (UK), ECMWF (UK) and Slovak Academy of Sciences (Slovak Republic). The {\it Boreasz} system
available at ICM, University of Warsaw is a single cabinet system with 2560 IBM Power7 compute cores and peak
performance of 78 TFlop/s. The main purpose of the system is research carried out within the POWIEW project \cite{powiew}
which among others includes scientific areas like large-scale cosmological simulations and numerical weather
forecasting. {\it Boreasz} is also part of the PRACE Tier-1 infrastructure \cite{prace}.


\section{Motivation}
\label{motivation}
Our motivation for developing an autotuning tool for OpenMP application was based on a specific behavior of chosen applications and synthetic benchmarks on the IBM Power 775 computational nodes. Measurements of our investigation for chosen codes and benchmarks are presented in this section. All of the benchmarks we have executed on the system were designed to examine the impact of using the multithreading mechanism to increase codes performance.

Firstly, we describe the effort we have made to measure performance of applications and synthetic benchmarks with the use of different simultaneous multithreading (SMT) modes. It should be stated that SMT mechanism does not increase the maximum number of FLOPs, however it might influence the performance of chosen algorithms and applications. This specific processor architecture feature is currently available in many petascale HPC systems available worldwide. Both IBM Power7 processors available in Power775 (IH) and IBM Power A2 processors available in Blue Gene/Q are built upon 4-way simultaneous multithreaded cores. It should be also mentioned that multithreading is predicted to be one of the leading features of future exascale systems available by the end of next decade \cite{dongarra}.

The first analysis of IBM Power 775 SMT mechanism was given in \cite{abeles}. It was shown that the performance gain from SMT varies depending on the program execution and its execution model, the threading mode being used on the processor, and the resource utilization of the program. The gains from using SMT modes with chosen algorithms where measured with the use of few well known
benchmarks: SPEC CFP2006, NAS Parallel Benchmark Class B (OpenMP) and NAS Parallel Benchmark Class C (MPI). One of
the conclusions of the study presented in [2] was that throughput type workloads are best suited to see gains from using higher
SMT modes. On the other hand high memory traffic codes will most likely not perform well when executed in SMT2 or SMT4
mode.
Through all of this paper we will extensively use the formulation that a specific application is using SMT2/SMT4 mode. By
saying this we will refer to parallel codes which are executed with number of processes and/or threads that exceed the physical
number of cores available in the system. This may be achieved by:
\begin{itemize}
\item executing an application with 2x or 4x more MPI processes,
\item executing an OpenMP/Pthreads code with 2x or 4x more threads,
\item mixing those two MPI and multithread execution modes (e.g. in the case of hybrid MPI/OpenMP codes).
\end{itemize}
Results presented in this section show that SMT mechanism available in modern processor chips can be efficiently use to increase performance of chosen applications and algorithms. On the other hand, there is a class of algorithms and applications
that does not benefit from mutithreading. In-depth investigation of the reasons of such divergence is needed and might be crucial for achieving good performance for complex computational codes.

\subsection{Performance analysis of scientific applications}

\subsubsection*{Computational cosmology.}

GADGET2 package was developed for large scale cosmological simulations on massively parallel computers with
distributed memory \cite{gadget}. It employs a variety of computational algorithms: hierarchical tree and particle mesh
algorithms for gravitational computations and Smoothed Particle Hydrodynamics (SPH) for modeling hydrodynamic of barionic
content of the Universe.

\begin{figure}[!h]
\centering
\includegraphics[scale=0.16]{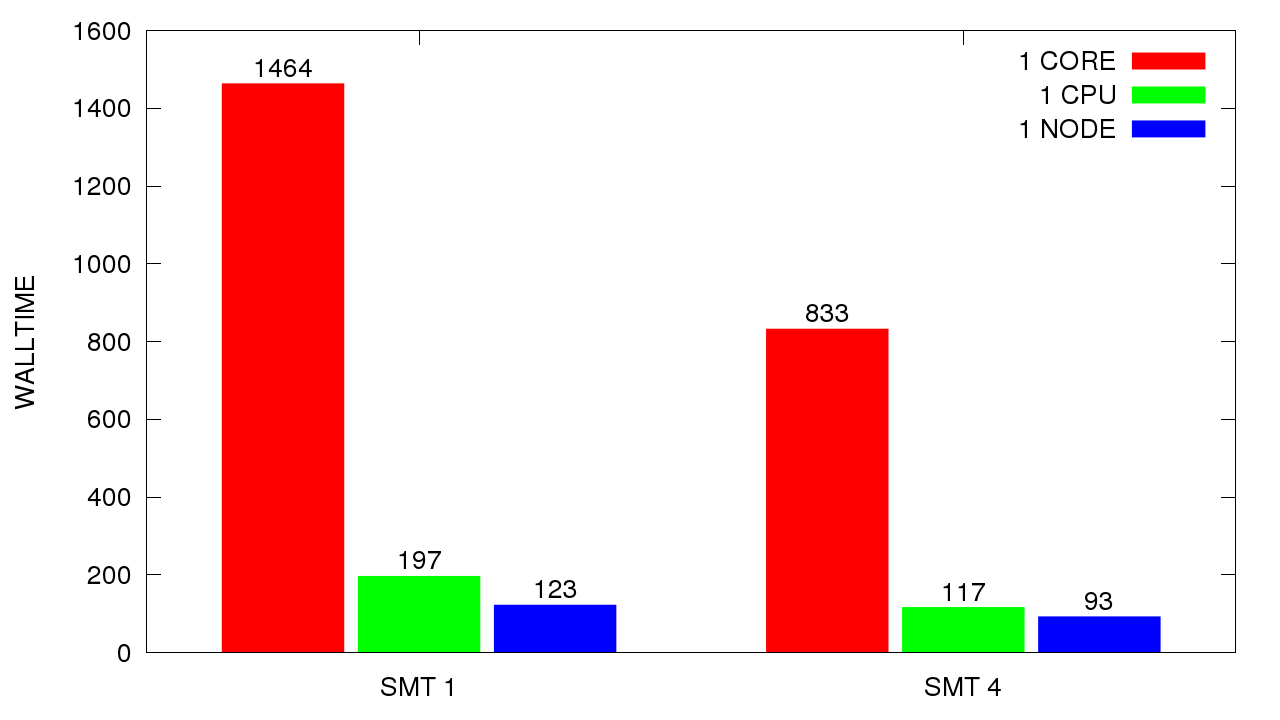}
\caption{Performance of GADGET2 application versus different SMT modes.}
\label{gadget}
\end{figure}

The test case used in this study consisted of
almost 28 million particles from which the half
were gas particles. The PMGRID parameter was
set to the size of 1024. The following compiler
options were used during the compilation phase on
the POWER7 system: {\tt -q64 -qarch=pwr7 -
qtune=pwr7 -O3 -qhot -qaltivec -qsimd=auto}.
Results of the measurements are shown in Figure \ref{gadget}.

\subsubsection*{Climate modeling.} 
WRF \cite{wrf} is a numerical weather prediction model
used for both operational forecasting and
atmospheric research in many leading
meteorological institutes around the world. WRF
code was prepared for execution on highly parallel computer architectures and is well known from its ability to scale using significant amount of processes and computing cores.

\begin{figure}[!h]
\centering
\includegraphics[scale=0.16]{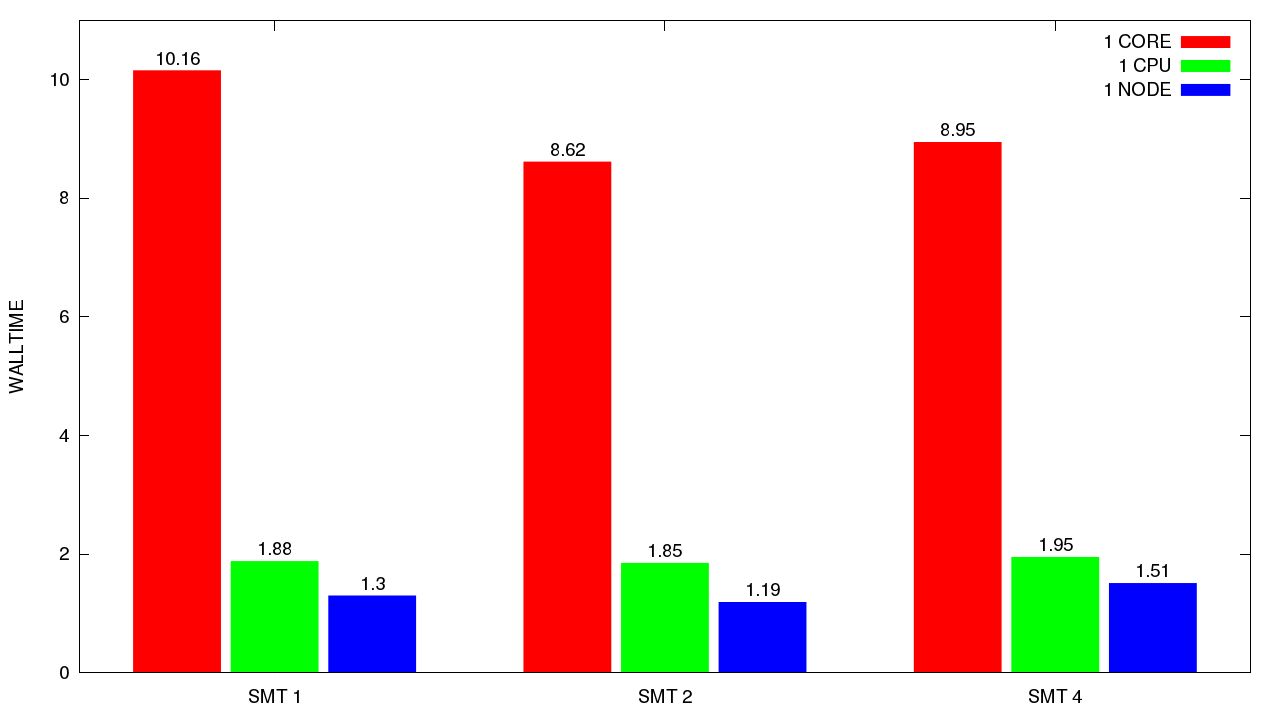}
\caption{Performance of WRF code measured against different SMT modes (mean elapsed time per time step).}
\label{wrf}
\end{figure}

For the purpose of this analysis we have used WRF v.3.2 compiled in the 64-bit mode and MPI-only version (DMPAR). We
have used January 2000 test case of WRF as a benchmark. The following compiler options were used during the compilation
phase on the POWER7 system: {\tt -q64 –O3 –qarch=pwr7 –qtune=pwr7 =qaltivec –lmass –lmassv}. Results of the measurements
are shown in Figure \ref{wrf}.

\subsubsection*{Molecular dynamics.}
We have selected two popular molecular dynamics
codes for our benchmark: GROMACS \cite{gromacs} and
CPMD \cite{cpmd}.

\begin{figure}[!h]
\centering
\includegraphics[scale=0.13]{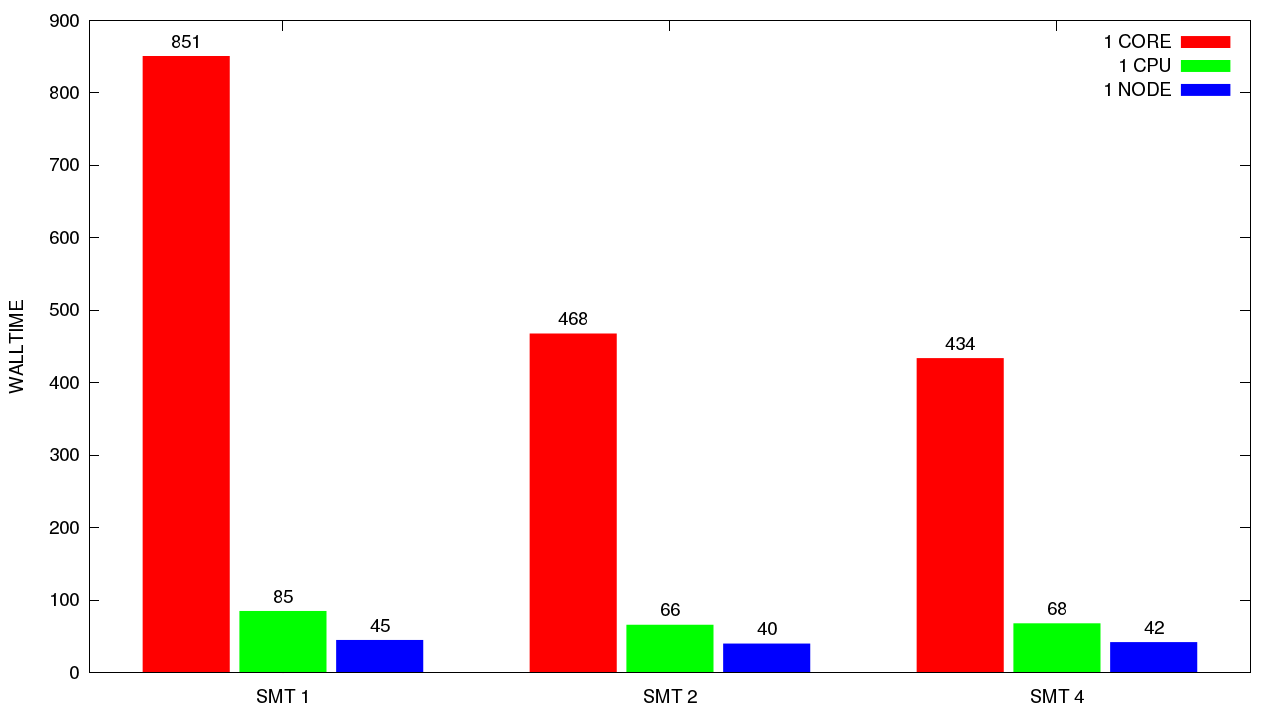}
\includegraphics[scale=0.13]{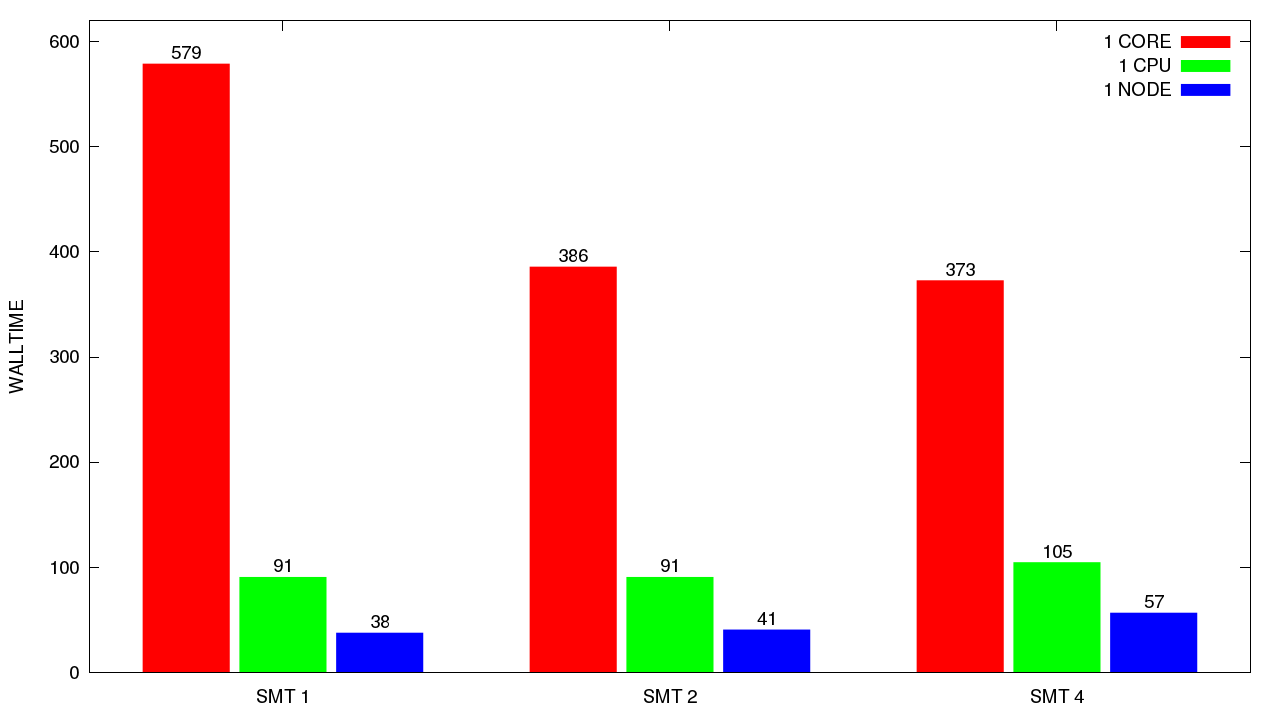}
\caption{Performance of GROMACS (left) and CPMD (right) codes measured against different SMT modes.}
\label{gromacscpmd}
\end{figure}

GROMACS is a package for performing molecular
dynamics simulations with hundreds to millions of
particles. It is implemented in C and Fortran and
uses MPI library for parallelization. The test case
used in this work was a vesicles in water system
consisting of more than 1 million atoms.
GROMACS v.4.0.7 was compiled and optimized
on Power7 system with the use of following
compiler options: {\tt -q64 –qarch=pwr7 –qtune=pwr7
–O3 –qhot –qalitvec –qsimd=auto}. Results of the
measurements are shown in Figure \ref{gromacscpmd}.

The CPMD code is an implementation of Density
Functional Theory (DFT), particularly designed for
ab-initio molecular dynamics. It runs on many
different computer architectures including parallel
systems. CPMD is parallelized with MPI and
OpenMP. End users can choose between different
versions of parallelization (distributed memory,
shared memory and mixed modes) during the
compilation phase. The test case used for
performance measurements was a water system
with 32 oxygen atoms and the mesh size of
128x128x128. CPMD v.3.13.2 was compiled and
optimized on Power7 system in the MPI-only
version with the use of following compiler options:
{\tt -q64 –qarch=pwr7 –qtune=pwr7 –O3 –qhot –qaltivec –qsimd=auto}. Results of the measurements
are shown in Figure \ref{gromacscpmd}. 

\subsubsection*{Materials sciences.}
\begin{figure}[!h]
\centering
\includegraphics[scale=0.16]{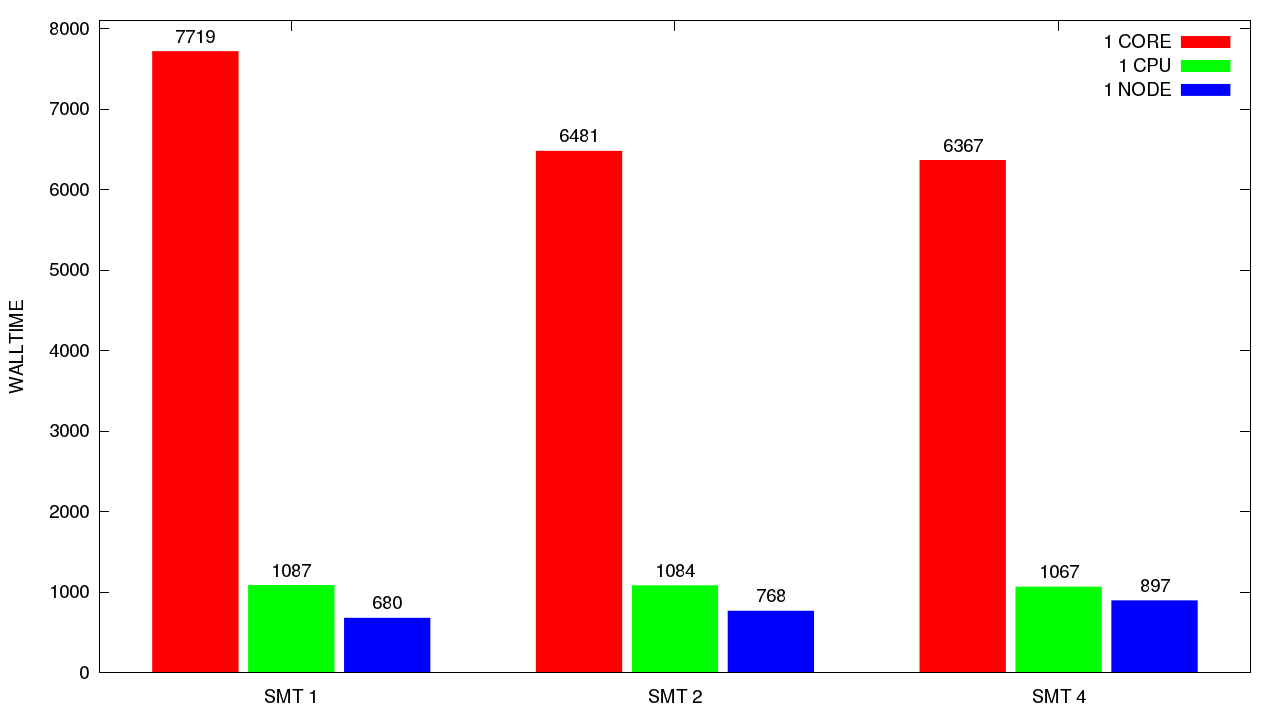}
\caption{Performance of GPAW code measured against different SMT modes.}
\label{gpaw}
\end{figure}

We have executed SMT performance tests with
GPAW simulation package \cite{gpaw}. GPAW is a
Density-Functional Theory (DFT) code based on
the projector-augmented wave method. It was
written in Python and C and requires NuMPy and
Atomic Simulation Environment (ASE) packages.
The parallel version of the code was prepared with
the use of MPI library. For the purpose of
performance analysis on Power7 system we have
run few iterations of a ground state calculations for
256 water molecules. GPAW v.0.7.6383 was
installed on the Power7 system together with its all
dependencies. The following options were used for C language parts of the package: {\tt -q64 –O3 –qhot –qaltivec –qarch=pwr7 –qtune=pwr7}. Results of the measurements are shown
in Figure \ref{gpaw}.

\subsubsection*{Conclusions}
As it was expected the performance gain from SMT varies when measured with different applications. GADGET2 is an example
of application which benefits from using SMT mechanisms. The best walltime results are always achieved for SMT4 mode
regardless of number of cores in use. On the other hand in the case of WRF and GPAW codes the expected performance gain
from using SMT is rather small (10\% in the case of WRF). The usage of SMT mechanism in the case of WRF seems to be
limited to SMT2. For GPAW the situation is even worst since using higher SMT modes decrease the overall performance.
Performance results obtained for GROMACS and CPMD show that those codes achieve rather good results from using SMT2
mode, but not from SMT4.
All test described above are also very important recommendations for users of the system since they can check which
type of SMT mode could be beneficial for their computations.

\subsection{Performance analysis of synthetic benchmarks}

Performance analysis of applications based only on walltime of the simulation run is sometimes not sufficient. To understand and identify performance bottlenecks an in-depth analysis of basic components of the applications (i.e. algorithms in use) is required. Therefore we have decided to analyze the performance of a representative set of algorithms which prove their usefulness in many scientific disciplines. Since the SMT mechanism is especially interesting in the case of multithreaded applications based on OpenMP/Pthreads model, we have decided to perform tests with BOTS benchmark \cite{bots}.
BOTS was developed as a basic set of applications that allows researchers and vendors to evaluate OpenMP implementations, and that can be easily ported to other programming models. An additional goal was for the OpenMP community to have a set of
examples using the tasking model available recently in OpenMP programming model.\\
Our testing procedure consisted of following steps:
\begin{enumerate}
\item Compile and execute BOTS benchmark on the Power775 system ({\it Boreasz}).
\item Identify the best scalable versions (tied/untied,for/single) through preliminary scalability testing of BOTS applications
on the single node of Power775 system.
\item Test the performance of selected versions of BOTS applications against different SMT modes.
\end{enumerate}
BOTS benchmark was compiled on {\it Boreasz} system with the use of following compiler options: {\tt -q64 –qalloca –qsmp=omp –O3
–qarch=pwr7 –qtune=pwr7 –qaltivec -qthreaded}.

The following versions of chosen applications have been identified:
\begin{itemize}
\item Strassen (tied tasks, manual - 3) – computes a matrix multiply with Strassen’s method, dense linear algebra algorithm
\item N Queens (tied tasks, manual -3) – finds solutions of the N Queens problem, search algorithm
\item SparseLU (tied tasks) – computes the LU factorization of a sparse matrix, sparse linear algebra algorithm
\item Health (tied tasks, manual - 2) – simulates a country health system, simulation algorithm
\item Floorplan (tied tasks, manual - 5) – computes the optimal placement of cells in a floorplan, optimization algorithm
\end{itemize}

Each of the application was executed with the largest possible input data provided within BOTS with the use of 32, 64 and 128
threads on the single node of Power775 system ({\it Boreasz}).

\begin{table}
\scriptsize
\renewcommand{\arraystretch}{1.3}
\caption{Performance of BOTS codes measured against different SMT modes.}
\label{bots_results}
\centering
\begin{tabular}{|l|c|c|c|c|}
\hline
\bfseries Application & \bfseries 1core,1thread & \bfseries 32 cores,32 threads & \bfseries 32 cores,64 threads & \bfseries 32 cores,128 threads\\
\hline \hline
\bfseries Strassen & 101.05s & 5.11s & 4.43s & 5.88s \\
\bfseries N Queens & 27.8s & 1.16s & 0.83s & 0.79s \\
\bfseries SparseLU & 129.19s & 4.79s & 4.18s & 5.03s \\
\bfseries Health & 96.54s & 3.97s & 3.7s & 4.8s \\
\bfseries Floorplan & 12.22s & 0.55s & 0.58s & 0.79s \\
\hline
\end{tabular}
\end{table}

As can be seen in the above Table \ref{bots_results} performance of Strassen, SparseLU and Health algorithms can benefit only from SMT 2 mode. N Queens search algorithm improve its performance also in the SMT 4 mode. Only for the Floorplan algorithm we did not see any increase of performance when using higher SMT modes. As it was pointed out before, we believe that precise
explanation of these results can be only given through analysis of chosen performance metrics and performance counters. 


\section{Towards Autotuning of OpenMP architectures}
\label{tools}

\subsection{Existing tools}

Code optimisation, despite rapidly increasing processor's power, is still very important part of developing scientific software.
A common solution to determine code's performance is to measure execution time of function calls or 
code's fragments that builds the core of the algorithm. This gives an important information where are the 
bottlenecks of the implemented algorithm and gives an idea of how much time is spent in selected parts of algorithm.
This is why many profiling tools where created that simplifies this process. Some of them, like gprof, measure
function calls execution times and trace call graph. Sometimes more detailed information is necessary.

Code instrumentation is a useful technique to measure code's performance. It allows to change the code automatically to
perform simple and more complicated tasks in order to find algorithm's properties as described above execution times.
IBM's hpctInst \cite{hpctInst} is a tool created by IBM for POWER systems that 
allows to instrument program's binary. 
It is a part of IBM's HPC toolkit which allows programmer to access hardware performance counters available
in POWER computers. Hardware performance counters give a detailed information about events, like L2 misses, missed branches and many more, that occurred during 
the execution of specified region together with time of execution.

HpctInst requires the binary to be compiled with -g option that includes into the binary
debugging information. It provides four types of instrumentation. One of them is 
an instrumentation for hardware performance counters, which is chosen by -dhpm flag.
It supports also instrumentation of MPI calls, OpenMP parallel constructs and IO operations, which are chosen by appropriate flags.

By default it instruments every entry and exit point of every function. 
Each profiling mode allows to provide a file that contains information about regions to instrument. 
Each line in a~file contains function name, filename and range of lines where the function calls should be instrumented. Only the function name is required. 
After the instrumentation is complete a~new binary is created that is ready to run.

When the instrumented binary has finished it's execution a file is created that contains the selected profiling information.
When environmental variable \texttt{HPM\_VIZ\_OUTPUT} is set to \texttt{yes} another file with viz extension is created.
The result file contains the profiling information in a human readable format, whereas viz file contains the same data in a xml format and 
can be opened with peekperf. 

If the code is instrumented for hardware performance counters then the result file contains number of selected events that occurred during the execution
of the selected regions. Instrumenting MPI calls results in gathering profiles for each process (one result file for one process) together with number of calls of 
each MPI function, time and average size of the message. Instrumenting parallel OpenMP constructs yields statistics about times of execution and for each thread.


\subsection{Design of the new tool}

Binary instrumentation that is available with hpctInst is a~useful tool but it has it's limits.
It allows only to instrument selected function calls and there are only four types of instrumentation. 
Moreover it is only available for xlc compiled binaries. 

A~tool developed by us, called PdtTagger, overcomes these shortages. PdtTagger is an instrumentation tool that operates 
on a~source code. So far it supports only code written in C. The tool is based on Program Database Toolkit (PDT) \cite{pdt} that provides access to a~high-level interface to source code. For a~given~source file it creates a~database file that contains information about the program structure.

Our tool basing on the database file and run-time options creates an~instrumented source file accordingly to given instructions. For selected regions it inserts 
library calls that indicates the beginning and the end of a~region. Inserting library calls allows to change the 
behavior during linking the application. The tool comes with a~library that allows to measure the time and hardware performance counters of selected regions
when working on POWER computers. Hardware counters are measured through interface provided by libhpm that is a~part of IBM's HPC toolkit.

By default our tool searches for OpenMP parallel constructs. Providing a~configuration file allows to select only selected regions of 
code for instrumentation. In that case it is a~programmers responsibility to ensure that the library calls are properly inserted into source code.
\begin{figure}[h!]
\includegraphics[scale=0.3]{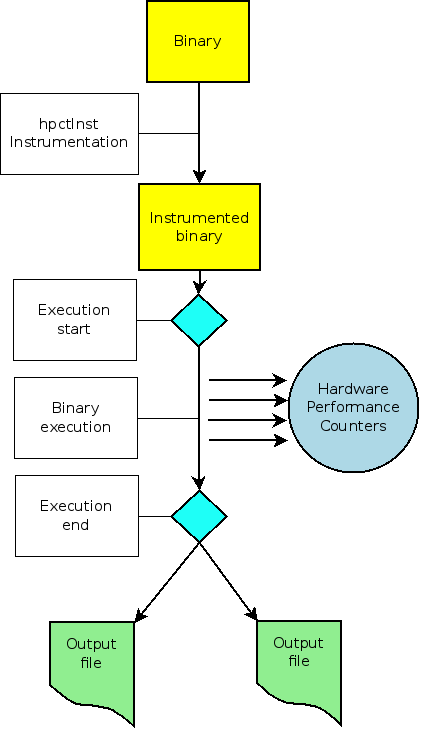}
\includegraphics[scale=0.3]{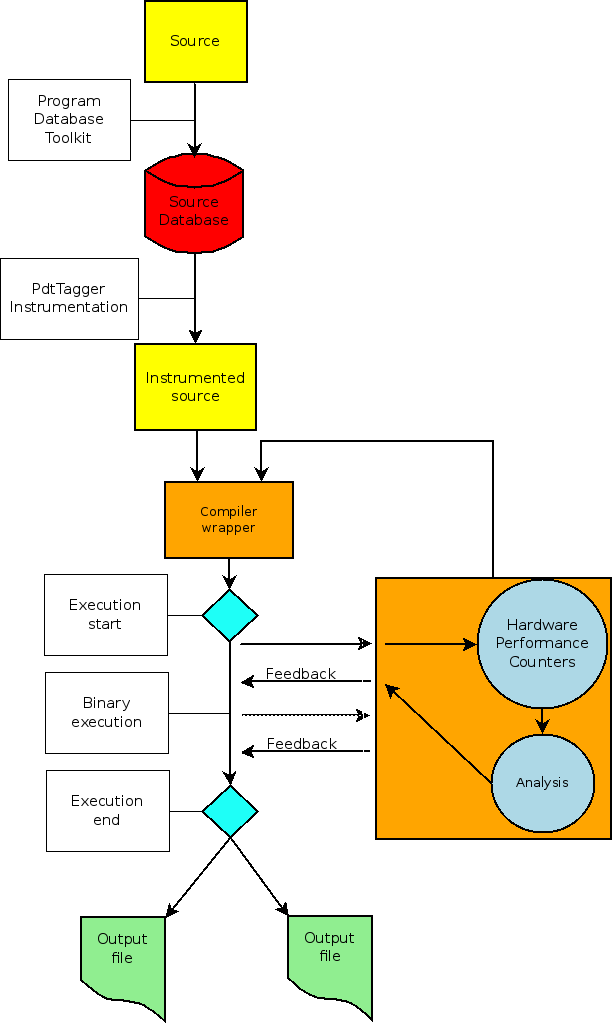}
\caption{Flowchart of hpcInst (left) and PdtTagger (right).}
\end{figure}

Compiling instrumented source is simple. It is sufficient to change the compiler in a~Makefile for PdtTagger compiler wrapper that links the application
with prepared library. When the instrumented source is compiled and the execution finished a~result file is created that contains execution times of selected
regions. If the application was linked with library that provides measuring hardware counters also the hpm file is created by libhpm library. 
As \texttt{HPM\_VIZ\_OUTPUT} affects libhpm it is possible to get the results also in viz format. It also possible
to create a~viz file for execution times by setting \texttt{PDTTAGGER\_VIZ\_OUTPUT} to \texttt{TRUE}.

The purpose for creating PdtTagger is to create a~tool that allows to automatically instrument the source code so that it selects 
appropriate number of threads that execute selected OpenMP~parallel regions. This could be achieved by linking the instrumented application with 
appropriate library. So far PdtTagger was successfully used to instrument Barcelona OpenMP~Task Suite and a~performance counters were gathered for
different types of applications. Constructing a~decision tree for a~selected representative set of counters could lead to library
basing on pmapi library, which allows to check performance counters at the run time, that will be able to suggest whether reducing or
increasing number of threads will speedup the execution of a~given region.


\section{Summary}

The most important advantages of the tool that has been described in this paper is that it provides an automatic code instrumentation of OpenMP parallel regions and it supports hybrid MPI and OpenMP applications. It was already tested for chosen applications from the BOTS benchmark. Further investigation of scientific codes is in progress. However, in order to construct the most appropriate automatic decision making mechanism we need to gather many results for different algorithms and computational kernels. The tool will be soon available for users of the Boreasz system.  


\section*{Acknowledgments}

This work was financially supported by the PRACE project funded in part by the EUs 7th Framework Programme (FP7/2007-
2013) under grant agreement no. FP7-261557. This research was carried out with the support of the "HPC Infrastructure for Grand Challenges of Science and Engineering" Project, co-financed by the European Regional Development Fund under the Innovative Economy Operational Programme. We would like to thank Maciej Marchwiany (ICM, University of Warsaw), Maciej Szpindler (ICM, University of Warsaw), Mateusz Wolszczak, Piotr Iwaniuk and Piotr Wojciechowski (MIM, University of Warsaw) for helping us to obtain sufficiently good performance with chosen benchmarks.

\bibliographystyle{abbrv}


\end{document}